\documentclass[reprint,nofootinbib,amsmath,amssymb,aps,preprintnumbers,superscriptaddress]{revtex4-1}

\usepackage{bm}
\usepackage{braket}
\usepackage{cases} 
\usepackage{color}
\usepackage{upgreek}
\usepackage{ulem}
\usepackage{comment}
\usepackage{empheq}

\usepackage{hyperref}
\hypersetup{colorlinks=true, allcolors=blue}

\newcommand{\ito}[1]{\textcolor{black}{#1}}

\begin{document}

\preprint{KEK-QUP-2023-0020, KEK-TH-2551, KEK-Cosmo-0322}

\title{Macroscopic Quantum Response to Gravitational Waves}

\author{Asuka Ito}
\email[]{asuka.ito@kek.jp}
\affiliation{International Center for Quantum-field Measurement Systems for Studies of the Universe and Particles (QUP), KEK, Tsukuba 305-0801, Japan}
\affiliation{
KEK Theory Center, Tsukuba 305-0801, Japan}

\author{Ryuichiro Kitano}
\email[]{ryuichiro.kitano@kek.jp}
\affiliation{KEK Theory Center, Tsukuba 305-0801, Japan}
\affiliation{Graduate University for Advanced Studies (Sokendai), Tsukuba 305-0801, Japan}


\begin{abstract}
We study the excitation of a one-electron quantum cyclotron by
gravitational waves. The electron in such as a penning trap is prepared to
be at the lowest Landau level, which has an infinite 
degeneracy parameterized by 
\ito{the spread of the wave function in position space.}
%
%
We find that the excitation rate from the ground state to the first
excited state is enhanced by the size of the electron wave function: an electron with a larger wave function
feels gravitational waves more.
As a consequence, we derive a good sensitivity to gravitational waves
at a macroscopic one-electron quantum cyclotron.
%
\end{abstract}

\maketitle

%
%
%
%
%
%
%

%
Gravitational waves are a powerful tool to explore our universe since
they bring valuable information as it has been demonstrated by
detection of gravitational waves around kHz with
interferometers~\cite{LIGOScientific:2016aoc} and of gravitational
waves around nHz with pulsar timing arrays~\cite{NANOGrav:2023gor}.
Certainly, advancing multi-frequency gravitational wave observations
is crucial for exploring our universe since different sources predict
different frequencies of gravitational waves~\cite{Kuroda:2015owv}. In
the lower frequency range of approximately $10^{-18}$ to $10^{-16}$
Hz, there is a viable method for observing gravitational waves,
namely, through the cosmic microwave background
observations~\cite{Planck:2018jri,Paoletti:2022anb,LiteBIRD:2022cnt}.
On the flip side, the detection of gravitational waves at frequencies
higher than kHz is still in the process of being developed, and it
even calls for novel ideas~\cite{Ito:2019wcb,Ito:2020wxi,Ejlli:2019bqj,Aggarwal:2020olq,Berlin:2021txa,Domcke:2022rgu,Tobar:2022pie,Berlin:2023grv,
Ito:2022rxn,Pshirkov:2009sf, Dolgov:2012be, Domcke:2020yzq, Ramazanov:2023nxz,Ito:2023fcr}. 
This is the case even though high frequency gravitational waves hold
theoretical interest for investigating new areas of
physics~\cite{Aggarwal:2020olq}.

One organic strategy to detect high frequency gravitational waves
involves employing small-scale experimental setups. This is attributed
to the fact that gravitational wave detectors tend to attain
heightened sensitivity when their dimensions align with the
wavelengths of the gravitational waves. For example, a magnon
gravitational wave detector has been proposed, which utilizes the
resonant excitation of magnons by gravitational waves to detect
signals around the GHz
range~\cite{Ito:2019wcb,Ito:2020wxi,Ito:2022rxn}. In the context of
this, there have been substantial proposals for pioneering methods of
detecting high frequency gravitational waves by leveraging experiments
designed for axion
detection~\cite{Ejlli:2019bqj,Aggarwal:2020olq,Berlin:2021txa,Domcke:2022rgu,Tobar:2022pie,Berlin:2023grv}.
However, further improvement of sensitivity is required to probe
theoretically interesting targets~\cite{Aggarwal:2020olq}. One
promising way to this end is employing quantum
sensing~\cite{RevModPhys.89.035002}.

In this letter, we propose the search for high frequency gravitational
waves by using a one-electron quantum cyclotron, which can also be
utilized for the dark photon dark matter search~\cite{Fan:2022uwu}. 
By recasting the reported dark count rate in a penning trap~\cite{Fan:2022uwu},
we give upper bounds on gravitational waves 
at frequencies of $O(100)$\,GHz, which have not been searched for. 
Generically, signal rates of gravitational wave detection are
suppressed by the ratio of the size of the detector and the wavelength 
of the gravitational wave, which is $O({\rm mm})$ in the case of
100~GHz wave. In this sense, a single electron does not sound a good
detector as it is small. 
However, our discussion reveals that the
state with a macroscopic wave function can compensate this factor.
In an ideal one-electron quantum cyclotron, each energy level has an
infinite number of degenerate energy states due to the translational (rotational)
invariance of the system. 
In the symmetric gauge where the Hamiltonian has the rotational
invariance, the degenerate states can be labeled by the size of the
wave function on the transverse plane.
We show that the excitation rate from the ground state to the first
excited state is enhanced when the ground state has a large wave
function.
In a realistic penning trap system where the degeneracy is lifted slightly,
the size of (the extent of) the wave function corresponds to that of a magnetron orbit.
Similarly, one also finds that the large axial motion enhances the
transition rate.
%
It is interesting to note that an electron with a larger wave function
feels gravitational waves more. The size of the wave function can be
macroscopic such as $O({\rm \mu m})\sim O({\rm mm})$ in the actual setup.
This fact may provide insights into the search for high frequency
gravitational waves using quantum sensing.
\section{One-electron quantum cyclotron} \label{Fermi}
In order to observe the magnetic moment of electrons, a one-electron
quantum cyclotron has been
utilized~\cite{Odom:2006zz,Hanneke:2008tm,Fan:2022eto}. Let us briefly
review the principle of the penning trap
(see~\cite{Brown:1985rh,Hanneke:2010au,Ito:2020xvp} for more details).
In a penning trap, an external magnetic field is applied on an
electron. Taking the direction of the magnetic field $B$ along the
$z$-direction, the electron experiences the cyclotron motion in the
$x$-$y$ plane and the spin precession. Furthermore, an external
\ito{inhomogeneous electric field $E$ is applied to bound the
electron in the $z$-direction}. The Hamiltonian of the system is given by
\begin{eqnarray}
  H &=& \frac{1}{2m_{e}} \left( p_{x} + \frac{eBy}{2} \right)^{2} + \frac{1}{2m_{e}}
        \left( \ito{p_{y}} - \frac{eBx}{2} \right)^{2} \nonumber \\
      & &+ \frac{1}{2m_{e}} p_{z}^{2} + \frac{1}{2}m_{e}\omega_{z}^{2}z^{2} - 2\mu_{B}S_{z} B \ ,  \label{hami}
\end{eqnarray}
where $m_{e}$ and $e$ are the mass and charge of an electron, $p_{i}$
represents the momentum, \ito{$\omega_z$ represents the angular frequency of the harmonic oscillator along $z$-axis}, $\mu_{B}$ is the Bohr magneton, and $S_{z}$
is the $z$ component of the spin operator. Note that we took the
symmetric gauge for the magnetic field, namely the vector potential
was taken to be $A_{i} = (-\frac{By}{2},\frac{Bx}{2},0)$, since this
gauge is the most convenient for the actual experimental setup, which
usually has a rotationally symmetric boundary. Also, the electric
field is identified as $E = -\frac{m_{e}\omega_{z}^{2}z}{e}$. We note
that although an electrostatic quadrupole potential is applied in the
penning trap actually, which affects the dynamics of the electron in
the $x$-$y$ plane, we will temporarily disregard the effect and
reassess it later. The $g$-factor of the electron has been set as
exactly $2$ for simplicity.

One can solve the time independent Schr\"{o}dinger equation with the Hamiltonian (\ref{hami}), 
and the following solution is obtained: 
\begin{equation}
  \Psi = \phi_{k}(z)  \psi_{n,m}(x,y) \chi_{\pm}\  ,
\end{equation}
where 
\begin{equation}
  \phi_k(z) = \frac{1}{\sqrt{2^{k}k!}} \left( \frac{m_{e}\omega_{z}}{\pi} \right)^{1/4} e^{-\frac{m_{e}\omega_{z}}{2}z^{2}}
            H_{k}\left( \sqrt{m_{e}\omega_{z}}z \right) \ ,
\end{equation}
%
%
is the wave function of the axial motion with $k=0,1,\dots \infty$
\ito{($H_{k}$ is the Hermite polynomial)}, and
\begin{eqnarray}
  \psi_{n,m}(x,y) &=& \frac{eB}{\sqrt{2\pi}} e^{im\varphi} \sqrt{  \frac{  (n-\frac{m+|m|}{2})!  }{  {(n-\frac{m-|m|}{2})!}  }  }
              e^{-\frac{eB}{4}\rho^{2}}    \nonumber \\
             & & \times \left( \frac{eB\rho^{2}}{2} \right)^{|m|/2}
                        L^{|m|}_{n-\frac{1}{2}(m+|m|)}\left(\frac{eB\rho^{2}}{2}\right) , \label{cyc}
\end{eqnarray}
with $n=0,1,\dots \infty$, $m = -\infty , \dots, n-1, n,$  is the wave
function of the cyclotron in the symmetric gauge in the $x$-$y$ plane
specified by the radial distance $\rho$ and the azimuth $\varphi$
\ito{($L^{m}_n$ is the generalized Laguerrel polynomial)},
and
\begin{equation}
  \chi_{+} = 
  \begin{pmatrix}
     1  \\ 0
  \end{pmatrix}  ,  \quad
  \chi_{-} = 
  \begin{pmatrix}
     0  \\ 1
  \end{pmatrix}  ,
\end{equation}
are the spinors for the spin up and the down states. The excitation of
the quantum cyclotron described by Eq.\,(\ref{cyc}) from the ground
state $(n=0)$ to the excited state $(n=1)$ can be measured with
penning traps~\cite{Odom:2006zz,Hanneke:2008tm,Fan:2022eto}. In
principle, we have infinite degenerate states labeled by $m\ (\leq n)$
for each energy level specified by $n$. However in practice, the
quantum number $m$ has a minimum value corresponding to the size of
the experimental apparatus because the size of the wave functions
become larger for larger $|m|$. Indeed, using the cyclotron frequency
$\omega_{c} = eB/m_{e}$, the root mean square of the radial distance
of the wave functions is given by $\sqrt{2(1-m)/m_{e}\omega_{c}}$ for
$(n=0, \ m\leq 0)$ and $\sqrt{2(3-m)/m_{e}\omega_{c}}$ for $(n=1, \
m\leq 1)$, and the wave functions exponentially decrease over them.
Therefore, the value of $m$ for these energy levels in a penning trap
with an radius $R$ must satisfies%
\footnote{ Since the wave functions exponentially decrease with a
decay constant $(m_{e}\omega_{c})^{-1/2}$ ($\ll R$) outside the root
mean square of the radial distance of the wave functions, the
definition of $m_{{\rm min}}$ is actually equivalent to require that
the wave functions sufficiently small at $R$. }
\begin{equation}
  |m| \lesssim m_{e}\omega_{c} R^{2} \sim 10^{9} \times \left( \frac{\omega_{c} /2\pi}{100 {\rm GHz}} \right) 
                                              \left( \frac{R}{0.5 {\rm mm}} \right)^{2}. \label{min}
\end{equation}
As we will see in the next section, this huge degeneracy can give rise
to enhancement of transition probability from the ground state to the
excited state by gravitational waves.

In the actual experimental setup of such as a penning trap, the degeneracy
for the $m$ quantum number is lifted by the electric field such that
each energy level has the interval of the magnetron frequency $\sim
\omega_z / 2 \omega_c^2$. 
In such a case, for a state with a large expectation value of $|m|$, it may be more
appropriate to take the coherent state.
The coherent state corresponds to the
classical magnetron motion with the radius 
$r_m \sim \sqrt{2 |m| / ( m_e \omega_c)}$. 
When $|m|$ is large, one can
understand $|m|$ as $ m_e \omega_c r_m^2 /2 $
by ignoring $O(1/|m|)$
quantities. 
Similarly, we usually have large $k$ for the axial motion in experiments,
and then the axial motion would also take the coherent state, i.e., $k \sim m_e \omega_z A^2 / 2$ with $A$
representing the amplitude of the oscillation in the axial direction.

In experiments~\cite{Odom:2006zz,Hanneke:2008tm,Fan:2022eto}, the
measurement of the excitation has been operated with the technique of
the quantum jump spectroscopy; one applies an additional weak magnetic
bottle field to couple the axial motion to the quantum cyclotron, so
that the energy transition of the quantum cyclotron can be measured
indirectly by monitoring the the axial frequency, which can be easily
measured by detecting the current induced by the axial motion.
Importantly, the interaction Hamiltonian commutes with the cyclotron
Hamiltonian, so that a quantum nondemolition measurement of the
cyclotron states is allowed. We note that since the spin precession
also couples to the axial motion due to the magnetic bottle field, one
can measure the spin excitation in the same way. However, we will see that
the spin excitation by gravitational waves is subdominant compared
with the excitation of the quantum cyclotron. In the following
section, we will review how to formulate the interaction between an
electron in a penning trap and gravitational waves.
\section{A Dirac particle in curved spacetime} \label{Fermi}
To investigate the effect of gravitational waves on a one-electron
quantum cyclotron, we consider the Dirac equation in curved spacetime
and take the non-relativistic limit assuming that the velocity of the
electron is much smaller than the speed of light as is satisfied in
real experiments. Furthermore, we need to use an appropriate
coordinate which comoves with the system.
Such a procedure has been carried out to study excitation of spins by 
gravitational waves~\cite{Ito:2020wxi} and/or to evaluate the inertial and gravitational effects due to the Earth on 
the magnetic moment of electrons/muons~\cite{Ito:2020xvp}.
More explicitly, the non-relativistic Hamiltonian for a Dirac particle such as an electron with mass $m_{e}$ 
in curved spacetime 
up to the order of $m_{e}^{0}$ is given by~\cite{Ito:2020xvp} 
\begin{eqnarray}
  H_{{\rm int}}  &=& \left( 1  + \frac{1}{2} R_{0k0l} x^{k} x^{l} \right)   m_{e}  
    - eA_{0}   \nonumber \\
   & &  + \frac{1}{3}  R_{0kil} \left( \Pi_{i} x^{k} x^{l}  
     +  x^{k} x^{l}  \Pi_{i} \right)  
     + \frac{1}{2} \epsilon_{0ijl} S^{l} R_{ijk0}  x^{k}  , \nonumber \\
      \label{Hint}
\end{eqnarray}
where $S^{i}$ stands for the spin operator, and $A_{\mu}$ represents
the vector potential of the electromagnetic field. The Riemann tensor
$R_{\mu\nu\rho\sigma}$ is evaluated at the origin of the coordinate
$x^{i}=0$, and $\Pi_{j} = -i \partial_{j} - eA_{j}$ is the canonical
momentum. The first term can be regarded as the rest mass and its
correction due to gravity. Compared with it, the third and the fourth
terms are higher order effects of $v/c \ll 1$ ($v$: speed of the
electron, $c$: speed of light) or $1/m_{e}|x^{i}| \ll 1$. Thus, we
will neglect them and focus on only the first term, which is at the
order of $m_{e}$. In particular, we see that there is no interaction
between the spin of the fermion and gravitational waves at the order.

We now consider gravitational waves on the Minkowski spacetime, the
metric is given by $g_{\mu\nu} = \eta_{\mu\nu} + h_{\mu\nu}$, where
$\eta_{\mu\nu}$ is the Minkowski metric and $h_{\mu\nu}$ is the
traceless transverse metric, namely, it represents gravitational
waves. Then, the Riemann tensor is 
\begin{equation}
    R^{\alpha}{}_{\mu\beta\nu} 
             =  \frac{1}{2}(h^{\alpha}_{\ \nu,\mu\beta}-h_{\mu\nu\ ,\beta}^{\ \ ,\alpha}
                   -h^{\alpha}_{\ \beta,\mu\nu}+h_{\mu\beta\ ,\nu}^{\ \ ,\alpha}) \ . \label{Rie}
\end{equation}
\ito{It should be noted that although we employs the transverse-traceless coordinate to calculate the Riemann tensor, the result is the same as that obtained in a Fermi normal coordinate since the Riemann tensor is gauge invariant at linear order.}
As a gravitational wave, which drives the excitation of the
one-electron quantum cyclotron, we consider a planer gravitational
wave:
\begin{eqnarray}
  h_{ij}(\bm{x}, t) &=& h^{(+)} \cos \left(\omega t - \bm{k}\cdot \bm{x}\right) e^{(+)}_{ij} \nonumber \\
           & & + h^{(\times)}\cos \left(\omega t - \bm{k}\cdot \bm{x} + \alpha \right) e^{(\times)}_{ij} \ , \label{planer}
\end{eqnarray}
where the polarization tensors satisfy $e^{(\sigma)}_{ij}
e^{(\sigma')}_{ij} = \delta_{\sigma\sigma'}$. $\alpha$ represents the
difference of the phases of the polarization. Note that the
polarization tensors can be explicitly constructed as
\begin{eqnarray}
  e _{ij}^{(+)} &=& \frac{1}{\sqrt{2}}\left(
    \begin{array}{ccc}
      \cos\theta^{2} & 0 & -\cos\theta \sin\theta \\
      0 & -1 & 0 \\
      -\cos\theta \sin\theta & 0 & \sin\theta^{2}
    \end{array} 
  \right) , \label{lipo1}  \\
  e _{ij}^{(\times)} &=& \frac{1}{\sqrt{2}}\left(
    \begin{array}{ccc}
      0 & \cos\theta & 0 \\
      \cos\theta & 0 & -\sin\theta \\
      0 & -\sin\theta & 0
    \end{array} 
  \right) .  \label{lipo2}
\end{eqnarray}
In the above Eqs.\,(\ref{lipo1}) and (\ref{lipo2}), we defined the $+$
mode as a deformation in the $y$-direction. Using
Eqs.\,(\ref{Rie})-(\ref{lipo2}) in Eq.\,(\ref{Hint}), we obtain the
interaction Hamiltonian between a one-electron quantum cyclotron and a
planner gravitational wave:
\begin{widetext}
\begin{equation}
  H_{{\rm int}} \simeq \frac{m_e\omega^{2}e^{i\omega t}}{8\sqrt{2}}
         \bigg[   \left( \cos^{2}\theta \, x^{2} + \sin^{2}\theta \, z^{2} -y^{2} - 2\cos\theta  \sin\theta  zx \right) h^{(+)} 
         + e^{i\alpha} \left(  2 \cos\theta  xy  
                - \sin\theta   yz \right) h^{(\times)}   \bigg] + ({\rm h.c.}).  \label{INT}
\end{equation}
\end{widetext}
In the next section, using the above interaction Hamiltonian, we will
calculate the excitation rate of a one-electron quantum cyclotron by a
planner gravitational wave and evaluate the ability of the penning
trap as a gravitational wave detector.
\section{Gravitational wave induced excitation} \label{GW} 
We are now ready to calculate the excitation rate of a one-electron
quantum cyclotron by gravitational waves. One can calculate the
excitation probability $P$ during an observation time $\tau$ at the
resonance point, i.e., the frequency of a gravitational wave coincides with
the cyclotron frequency, by using
Eq.\,(\ref{INT})~\cite{griffiths:quantum}:
\begin{equation}
  P \simeq \sin^{2} \left( \frac{\left|  \braket{\Psi_{{\rm out}} | \tilde{H}_{{\rm int}} | 
                       \Psi_{{\rm in}}} \right|}{2} \tau \right) , \label{rate}
\end{equation}
where $\tilde{H}_{{\rm int}}$ is defined by $H_{{\rm int}} =
\tilde{H}_{{\rm int}} e^{i\omega t} + ({\rm h.c.})$. Note that we
implicitly assumed that the gravitational wave has a coherence during
the observation time. More explicitly, from Eqs.\,(\ref{cyc}) and
(\ref{INT}), one can calculate the matrix element in Eq.~\eqref{rate}
for the transition from a ground state to an excited state as
\begin{widetext}
\begin{subequations}
    \begin{empheq}[left = {\empheqlbrace \,}, right = {}]{align}
     \left| \braket{1,m+2,k | \tilde{H}_{{\rm int}} | 0,m,k} \right|^{2} &=  
        \frac{|m|\omega^{2}}{128}\left[ \left( (1+\cos^{2}\theta) h^{(+)} - 2\cos\theta\sin\alpha h^{(\times)} \right)^{2} + 
                                        4 \cos^{2}\theta \cos^{2}\alpha (h^{(\times)})^{2} \right], \label{am0} \\
     \left| \braket{1,m,k| \tilde{H}_{{\rm int}} | 0,m,k} \right|^{2} &=
       \frac{(|m|+1)\omega^{2}}{128}  \sin^{4}\theta (h^{(+)})^{2}  , \label{am} \\
     \left| \braket{1,m+1,k+1 | \tilde{H}_{{\rm int}} | 0,m,k} \right|^{2} &=
        \frac{\left(k+1\right) \omega^{3}}{32\omega_{z}} \sin^{2}\theta 
        \left[ \left( \cos\theta h^{(+)} - \sin\alpha h^{(\times)} \right)^{2} + 
                                        \cos^{2}\alpha (h^{(\times)})^{2} \right]   ,  \label{am2} \\
     \left| \braket{1,m+1,k-1 | \tilde{H}_{{\rm int}} | 0,m,k} \right|^{2} &=
        \frac{k \omega^{3}}{32\omega_{z}} \sin^{2}\theta 
        \left[ \left( \cos\theta h^{(+)} - \sin\alpha h^{(\times)} \right)^{2} + 
                                        \cos^{2}\alpha (h^{(\times)})^{2} \right] .  \label{am3}
    \end{empheq}
\end{subequations}
\end{widetext}
Eqs.\,(\ref{am0}) and (\ref{am}) come from $\tilde{H}_{{\rm int}}$ including only $x$
and/or $y$ components. It is worth noting that the transition
probability increases for larger $|m|$. This is a sort of particular
feature of the excitation by gravitational waves in contrast to the
excitation caused by electromagnetic fields; we have the same
transition probability from $\ket{0,m}$ to $\ket{1,m+1}$ for arbitrary
$m$ in the case \ito{of dipole excitation by electric fields.}
Interestingly, it enables us to increase the
sensitivity of a one-electron quantum cyclotron to gravitational waves
by 
enlarging the wave function, whose size is proportional to
$\sqrt{|m|}$. On the other hand, Eqs.\,(\ref{am2}) and (\ref{am3}) are calculated from
$\tilde{H}_{{\rm int}}$ including $z$ component. It also gets increased for
larger $k$. Thus, we can also increase the
transition rate by preparing a large wave function in the
$z$-direction.

Now let us evaluate the sensitivity of a one-electron quantum
cyclotron to a planer gravitational wave by using
Eqs.\,(\ref{rate})-(\ref{am3}). Supposing that we have an initial
state $\ket{\Psi_{{\rm in}}} = \ket{0,m,k}$, it can be excited to
$\ket{1,m+2,k}$ through Eq.\,(\ref{am0}), $\ket{1,m,k}$ through Eq.\,(\ref{am}), or
$\ket{1,m+1,k\pm 1}$ through Eqs.\,(\ref{am2}) and (\ref{am3}). Since the transition rate
depends on $m$ and $k$, we adopt their values used in
experiments~\cite{Odom:2006zz,Hanneke:2008tm,Fan:2022eto} as a
reference. 
We then need to look back on Eq.\,(\ref{hami}), in which we
just considered an electric field along the axial direction. 
However, in real measurements, a quadruple potential is used to combine an
electron in the axial axis, so that the cyclotron motion in the
$x$-$y$ plane is also modified by the potential. 
Then the degeneracy
of the cyclotron energy levels are lifted due to the appearance of
the magnetron motion~\cite{Brown:1985rh,Hanneke:2010au,Ito:2020xvp}.
A specific value of $m$
would be chosen in such a measurement. 
Indeed, $|m| \sim k \sim
10^{3}$ is obtained in the cooling limit with a sideband
cooling~\cite{Brown:1985rh}. 
As we already mentioned, when the quantum numbers $|m|$ and $k$ are much larger than unity,
coherent states are realized for the magnetron motion and the axial motion, respectively.
However, it does not change the evaluation of the excitation of a one-electron quantum cyclotron by 
gravitational waves \ito{at least approximately, since the coherent state has a Gaussian distribution with a peak at $|m|\sim k \sim 10^{3}$}.
Using the upper limit on the dark count
rate per unit time, $4.33 \times 10^{-6}$ s$^{-1}$, obtained
in~\cite{Fan:2022uwu}, one can evaluate the sensitivity of a
one-electron quantum cyclotron to a planner gravitational wave through
the transition of (\ref{am0}) and (\ref{am}),
\begin{equation}
  h_{0} \sim 1.2 \times 10^{-18} \left( \frac{\omega/2\pi}{200 {\rm GHz}} \right)^{-1} 
                                 \left( \frac{|m|}{1000} \right)^{-1/2}   
                                 \left( \frac{\tau}{1{\rm day}} \right)^{-1/2}  , \label{sens}
\end{equation}
and of (\ref{am2}) and (\ref{am3}),
\begin{eqnarray}
  h_{0} \sim 3.8 \times 10^{-20} & & \left( \frac{\omega/2\pi}{200 {\rm GHz}} \right)^{-3/2} 
                                 \left( \frac{\omega_{z}/2\pi}{100 {\rm MHz}} \right)^{1/2} \nonumber \\
                    & &\times             \left( \frac{k}{1000} \right)^{-1/2}   
                                 \left( \frac{\tau}{1{\rm day}} \right)^{-1/2}  . \label{sensz}
\end{eqnarray}
In the estimation, we took average of $\theta$ because we do not know
the value in advance and assumed no polarization by setting $\alpha =
0$, and $h^{(+)} = h^{(\times)} = h_{0}$. 
Eqs.\,(\ref{sens}) and (\ref{sensz}) would be valid
for a frequency range from $20$ GHz to $200$ GHz because the cyclotron
frequency can be tuned in the range, as discussed
in~\cite{Fan:2022uwu}. 
\ito{We took an observation time of 1 day just as a reference value.
Note that it may significantly deviate from the coherence time of expected gravitational wave sources~\cite{Aggarwal:2020olq}}.

Finally, let us consider a strategy to improve the sensitivity to
gravitational waves by preparing a macroscopic one-electron quantum
cyclotron. 
As Eqs.\,(\ref{am0})-(\ref{am3}) tell us, the transition rate by
gravitational waves is proportional to the size of (the extent of) the wave function,
which is parameterized by $|m|$ and $k$. 
One can increase $|m|$ and $k$ arbitrarily in principle. 
However, in practice, there would be a maximum value for them like Eq.\,(\ref{min})
specified by the size of the apparatus of a penning trap. 
Supposing that we can tune the value of $|m|$ to be the maximum value,
$m_{e}\omega_{c}R^{2}$, and $k$ is also taken to be $k\sim |m|$ by a sideband cooling,
the sensitivity of a one-electron quantum
cyclotron to a gravitational wave would be
\begin{equation}
  h_{0} \sim 2.2 \times 10^{-21} \left( \frac{\omega/2\pi}{200 {\rm GHz}} \right)^{-3/2} 
                                 \left( \frac{R}{0.2{\rm mm}} \right)^{-1}   
                                 \left( \frac{\tau}{1{\rm day}} \right)^{-1/2}  , \label{sens2}
\end{equation}
through the transition (\ref{am0}) and (\ref{am}), and
\begin{eqnarray}
  h_{0} \sim 6.9 \times 10^{-23} & &\left( \frac{\omega/2\pi}{200 {\rm GHz}} \right)^{-2} 
                                 \left( \frac{\omega_{z}/2\pi}{100 {\rm MHz}} \right)^{1/2} \nonumber \\
                         & &\times         \left( \frac{R}{0.2{\rm mm}} \right)^{-1}   
                                 \left( \frac{\tau}{1{\rm day}} \right)^{-1/2}  , \label{sens3}
\end{eqnarray}
through the transition (\ref{am2}) and (\ref{am3}).
As expected, we see that the sensitivity is increased for larger $R$
because then the wave function becomes more macroscopic. Thus, we can
improve the sensitivity by preparing a large apparatus. However, we
note that one can not take arbitrary large value of $R$ in our
formalism since 
$\omega R \lesssim 1$ should be satisfied
to validate the interaction Hamiltonian (\ref{Hint}). It is because
that Eq.\,(\ref{Hint}) has been derived by truncating the infinite sum
of terms in the metric of Fermi normal coordinates by assuming
$\omega R \lesssim 1$~\cite{Ito:2020wxi,Ito:2020xvp}. If
the assumption does not hold, one needs to take into account infinite
terms in the metric as was done in~\cite{Ito:2022rxn}. As a
consequence, it is natural to expect that the sensitivity would have a
peak around $\omega R \sim 1$ at which the wavelength of
the gravitational wave is comparable to the size of the wave
function~\cite{Ito:2022rxn}. In the evaluation (\ref{sens}), we took
$R = 0.2{\rm mm}$ to satisfy $\omega R \lesssim 1$ at
$\omega/2\pi = 200$GHz. 
Therefore, for example, we can take $R=2$mm when $\omega/2\pi = 20$GHz is considered.
Notably, such a setup can be realized in
the existing
experiments~\cite{Odom:2006zz,Hanneke:2008tm,Fan:2022eto}.
\section{Conclusion} \label{con}
We studied the excitation of a one-electron quantum cyclotron by
gravitational waves. We calculated the transition rate from degenerate
ground states, in which each wave function has a different size, to
first excited states. It turned out that the transition rate from the
ground state to the first excited state is enhanced when the ground
state has a larger wave function. As a consequence, good sensitivities
to high frequency gravitational waves are derived for a
macroscopic one-electron quantum cyclotron.
Therefore, in principle, one can increase the sensitivity to
gravitational waves just simply preparing a ground state with a larger
wave function. 
This fact may shed light on searches for high frequency
gravitational waves by quantum sensing.
There is a related study in Ref.~\cite{Fischer:1994ed} where it is found that gravitational wave signals are enhanced by using an atom in a highly excited state such as a Rydberg atom. 
In contrast to atoms, the penning trap experiment does not require preparing such a special state, and the enhancement is more or less automatic.
Another virtue of the experiment is the possibility to scan the
frequencies of probed gravitational waves by changing the strength of the magnetic field and the
frequencies of driving fields. Indeed, one can search for
gravitational waves within a frequency range of $20$ to $200$GHz with
a penning trap, maintaining the sensitivity~\cite{Fan:2022uwu}.
\begin{acknowledgments}
A.\,I.\ was supported by World Premier International Research
Center Initiative (WPI), MEXT, Japan. This work is supported by JSPS KAKENHI Grant Numbers
JP22K14034~(A.I.),
JP22K21350~(R.K.),
JP21H01086~(R.K.), and
JP19H00689~(R.K.).
\end{acknowledgments}

\bibliography{ref}

\end{document}